\documentclass{article}

\usepackage{graphicx} 
\usepackage{amsmath}
\usepackage{bbm}

\usepackage[dblblindworkshop, final]{neurips_2026}
\workshoptitle{GEM}

\makeatother

\usepackage[utf8]{inputenc} 
\usepackage[T1]{fontenc}    
\usepackage{hyperref}       
\usepackage{url}            
\usepackage{booktabs}       
\usepackage{amsfonts}       
\usepackage{nicefrac}       
\usepackage{microtype}      
\usepackage{xcolor}         
\usepackage{enumitem}
\usepackage[most]{tcolorbox}
\usepackage{fontawesome5}   
\usepackage{xcolor}

\definecolor{cardaccent}{HTML}{7A5BC0}   
\definecolor{purple}{HTML}{5A3E9B}   
\definecolor{blue}{HTML}{185FA5}     
\definecolor{purple2}{HTML}{7A5BC0}  
\definecolor{teal}{HTML}{0F6E56}     

\newtcolorbox{taskcard}[1]{
  breakable,
  enhanced jigsaw,
  colback=cardaccent!6, colframe=cardaccent,
  coltitle=white, fonttitle=\bfseries\sffamily\small,
  title={\faDna\ \ #1},
  boxrule=0.8pt, arc=3pt,
  left=10pt, right=10pt, top=8pt, bottom=8pt,
  before skip=8pt, after skip=8pt,
}
\newcommand{\cardtop}[1]{\noindent\textbf{\footnotesize #1}\par}

\newlist{mh}{enumerate}{1}
\setlist[mh]{label=\arabic*., leftmargin=1.6em, itemsep=2pt, topsep=3pt, parsep=0pt}
\newlist{sg}{itemize}{1}
\setlist[sg]{label=\textendash, leftmargin=1.4em, itemsep=2pt, topsep=3pt, parsep=0pt}

\newtcolorbox{addonnote}[1]{
  colback=gray!8, colframe=gray!40,
  boxrule=0.5pt, arc=2pt,
  fontupper=\small,
  left=8pt, right=8pt, top=6pt, bottom=6pt,
  before skip=6pt, after skip=6pt,
  before upper={\textbf{\footnotesize #1}\par\vspace{2pt}},
}

\tcbuselibrary{breakable,skins,listings}

\lstdefinestyle{promptstyle}{
  basicstyle=\ttfamily\scriptsize,
  breaklines=true, breakindent=0pt, breakatwhitespace=false,
  columns=fullflexible, keepspaces=true, showstringspaces=false,
  frame=none, aboveskip=0pt, belowskip=0pt,
}

\newtcblisting{promptbox}[1]{%
  breakable,
  colback=black!3, colframe=black!45, boxrule=0.5pt, arc=2pt,
  left=4pt, right=4pt, top=4pt, bottom=4pt,
  title={#1},
  fonttitle=\bfseries\small, coltitle=black, colbacktitle=black!10,
  listing only, listing options={style=promptstyle},
}
\title{Can agents reliably design therapeutic antibodies?}

\author{%
  Asli Alpman\thanks{Work done while at BigHat Biosciences} \\
  UC Berkeley\\
  \texttt{asli.alpman@berkeley.edu} \\
  \And
  Nick Bhattacharya\\
  BigHat Biosciences
  \And
  Aaron Solomon\\
  BigHat Biosciences
  \And
  Hunter Elliott\\
  BigHat Biosciences
}

\begin{document}

\maketitle

 \begin{abstract}
 Recent work has demonstrated agents driving protein design workflows, either through orchestration of specialized tools or direct manipulation of sequences. Yet the fidelity with which agents follow protein design instructions and the factors contributing to their effectiveness remain understudied. We examine agentic antibody design under two settings: when explicit molecular-editing instructions are provided and when agents receive only previously-scored sequences and are prompted to propose improvements. We call the first task-guided and the second black-box design. For the task-guided setting, we curate ten expert-validated affinity-improvement tasks and show that LLM judges closely reproduce programmatic evaluations of instruction following. We then study two augmentations: access to antibody-relevant tools and to previously scored design history. Tools improve instruction following (IF) and sequence space exploration while reducing optimization performance, whereas design history reduces IF but improves optimization performance at the expense of exploration. Finally, we combine task-guided and black-box design within an in silico optimization loop and show that our agentic optimizer performs comparably to an established Bayesian optimization baseline.
\end{abstract}

\section{Introduction}

Large language models (LLMs) are increasingly deployed as agents in scientific discovery, coupling natural-language reasoning to domain-specific tools to plan and execute multi-step workflows across chemistry, materials, and the life sciences \citep{lu2024aiscientist,zheng2025automation,panigrahi2026heurekabench,teneggi2026agentrosetta}. This framework fits well with protein engineering, where iterative generate-and-evaluate workflows align naturally with agents that plan, call tools, and refine their outputs \citep{kim2026biodesignbench}. In protein engineering, agents have been used both as optimizers that propose and refine candidate designs \citep{wang2025large, wang2025swarms, maus2026pablo} and as orchestrators that coordinate tools \citep{ghafarollahi2024protagents, jacob2025geniecat, ge2026protocycle, xu2026proteinmcp}. Yet as capabilities grow, a more basic question is easy to overlook: when a scientific agent is given a specific protein design instruction, does it do what it was told, and what determines whether it does?

Here we study iterative antibody design in which agents propose targeted sequence edits, and jointly evaluate adherence to design instructions and affinity improvement \textit{in-silico}, unlike existing benchmarks that evaluate either protein design performance \citep{kim2026biodesignbench} or instruction following in text-based scientific QA \citep{su2026sciif} independently. We consider two modes of agentic design: a \textit{black-box} setting, in which agents condition only on previously characterized sequences, and a \textit{task} setting, in which agents receive explicit edit instructions. We systematically ablate access to tools and characterized design history across six models from the Claude and GPT families. We then combine black-box and task agents in a multi-agent iterative design loop that achieves performance comparable to a previously published Bayesian optimization method. 

\section{Methods}

\subsection{Task and black-box setting}
We curate a set of 10 tasks aimed at improving affinity, each validated by a domain expert. Each task is specified through a design brief including a summary, a set of must-have requirements, and a set of optional suggestions. The must-have requirements are stated as mandatory constraints and are used to evaluate instruction following, while the suggestions are presented as optional guidance. Details of all tasks are provided in Appendix~\ref{appendix:design-tasks}.  We also consider a \emph{black-box} setting in which the agent is given only prior sequences and their scores and is prompted to use the design history to propose improvements. 

\subsection{Agent configuration}\label{sec:agentconfig}

\textbf{Models.} We evaluate frontier models from the Claude and GPT families, including Claude Opus 5, Opus 4.8, and Sonnet 5, as well as GPT-5.6 Sol, GPT-5.6 Terra, and GPT-5.6 Luna. All models are evaluated at their highest available reasoning setting. For Claude models with high refusal rates, we instead use the next-lower reasoning setting, which we empirically observe to reduce refusals. Sonnet 5 is used only as an LLM judge, as it refuses nearly all calls when evaluated as a task-performing agent. Prompt configurations and full prompt details are provided in Appendix~\ref{appendix:prompts}.

\textbf{Tools.} Agents are equipped with a suite of established tools for antibody design, which they may invoke at their discretion. These include ANARCI~\citep{dunbar2016anarci} for IMGT numbering; NanoBodyBuilder2~\citep{abanades2023immunebuilder} for fold prediction; FreeSASA~\citep{mitternacht2016freesasa} for per-residue solvent-accessible surface area (SASA); nanoBERT~\citep{hadsund2024nanobert}, a nanobody-specific protein language model; and Biopython~\citep{cock2009biopython} for isoelectric point (pI) computation and other basic sequence utilities such as residue counting.


\subsection{Instruction following evaluation}\label{sec:IF}
Each agent invocation maps a (task, seed) pair to a batch of candidate output sequences. Each task has a set of must-have instructions and instruction following is measured based on these must-haves. To determine whether must-haves are satisfied, we employ two approaches. First, we have programmatic verifiers which run deterministic code on agent output. Tasks are curated such that these are available. Second, we have LLM judges which are prompted to generate pass/fail for each must-have in a (task, seed, output) triplet. We quantify instruction following with two metrics, \emph{strict compliance} and \emph{partial compliance}. Strict compliance is a binary metric which is 1 if and only if all must-haves are satisfied. Partial compliance is the average of all individual must-have compliances. 



\subsection{Design loop}
We embed the tasks in an iterative agentic design loop driven by a simple surrogate. 
From the ten design tasks we select the four most effective (Charge Repair, Surface Targeted, Backbone Shape, Paratope Electrostatics), along with the black-box task. All
evaluation is done in-silico using an antibody affinity oracle from
\citet{ober2025sequence}. This is an in-house model trained on data from a real-world
optimization campaign; the design objective is improvement in the antibody's
predicted dissociation constant $K_D$ relative to the parental. The Bayesopt method from \citet{ober2025sequence} as well as PABLO \citet{maus2026pablo}, an agentic black box optimizer, are used as baselines (see Appendix~\ref{appendix:configs} for details). 
We start all methods from 50 seed sequences drawn from the early stages of that campaign and proceed for 10
rounds, acquiring 50 molecules per round for a total oracle budget of 500. To understand how each component of our multi-agent system contributes, we also ablate the black-box agent and the task agents separately. Lastly, we analyze the positions where mutations are introduced by all three methods. Recall that antibodies have framework regions (FRs) and CDR regions; CDRs are key drivers of binding to a specific target, while FRs typically play a smaller role in binding and are key for many other key properties. Accordingly, we pay special attention to the proportion of mutations made in FRs versus CDRs (using IMGT numbering). 

\section{Results}

\begin{figure}[h]
  \label{fig:judge-toolefficieny}
  \centering
  \includegraphics[width=0.4\textwidth]{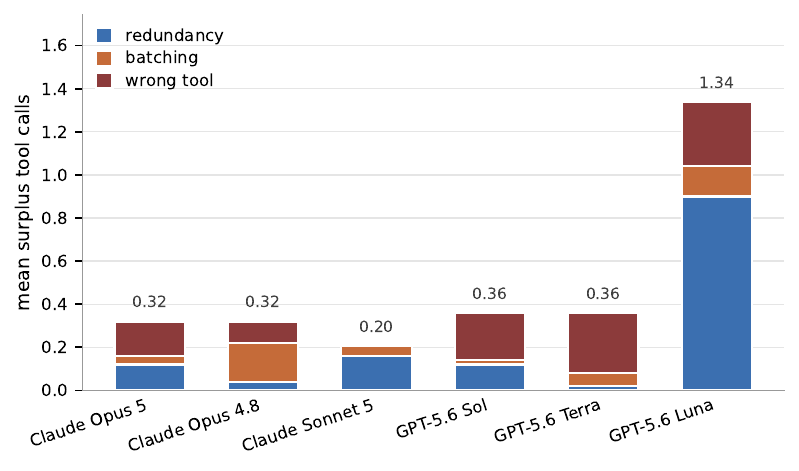}
\caption{\textbf{Surplus tool calls by LLM judges.} Taking the number of 
tool calls required by the programmatic checks as the reference minimum; surplus 
is the number of issued calls above this minimum, averaged over tasks. We 
decompose surplus calls into \textit{redundancy} (re-fetching a 
result already in hand), \textit{batching} (issuing combinable calls separately 
instead of in one call), and \textit{wrong tool} (calling a tool the task did 
not require).}
\end{figure}

\paragraph{LLM judges agree with programmatic verifiers.}
We first ask whether LLM judges, given the same toolset as in \ref{sec:agentconfig}, can reproduce the programmatic labels on the same outputs (Table~\ref{tab:judge-agreement}). Among the judges that returned verdicts, agreement is near-perfect for every model except GPT-5.6-luna. 
Every must-have here is defined as a property the tools can measure, so we interpret this as follows: for tool-relevant constraints, a judge's verdicts line up with deterministic checks. Using the number of tool calls required by the programmatic checks as a reference, we compute the surplus tool calls made by LLM judges. The mean surplus is small for most judges
(see Figure~\ref{fig:judge-toolefficieny}) and the dominant source of surplus
differs by model. 

\begin{table}[h]
\centering
\footnotesize
\caption{LLM-judge agreement with the programmatic checks over 1{,}900
(output, must-have) cells per judge: 50 (task, seed) pairs (10 tasks
$\times$ 5 seeds), each holding 10 outputs scored against 3--4 must-haves.
We report recall and precision with FAIL (a violation) as the positive
class; refused cells yield no verdicts and are excluded from the analysis.}
\label{tab:judge-agreement}
\begin{tabular}{lrrr}
\toprule
Judge & Recall (\%) & Precision (\%) & Refusal rate (\%) \\
\midrule
Claude Opus 5    & 100.0 & 100.0 & 0.0 \\
Claude Opus 4.8  & 100.0 & 100.0 & 0.0 \\
Claude Sonnet 5  & 100.0 & 100.0 & 12.0 \\
GPT-5.6 Sol      & 100.0 & 100.0 & 0.0 \\
GPT-5.6 Terra    & 100.0 & 100.0 & 0.0 \\
GPT-5.6 Luna     & 100.0 & 94.6  & 0.0 \\
\bottomrule
\end{tabular}
\end{table}

\paragraph{Instruction following failures are common, but tool access resolves them.}
Every model has a gap between partial and strict compliance when no tools are available (see~Table~\ref{tab:if-pass-rate}). These failures are rarely complete failures; models usually follow most must-haves. When prior history is added, the strict compliance falls even further relative to partial compliance (see~Table~\ref{tab:if-pass-rate}). The effect is not
family specific, with the two largest drops at Claude Opus~4.8 (0.31) and GPT-5.6-terra (0.23). 
Details on tool usage are provided in Appendix~\ref{appendix:toolUsage}. See Section \ref{appendix:if-failure} for a partial categorization of instruction following errors.

\begin{table*}
\caption{\textbf{Tool access improves instruction following.} Strict and
partial compliance rates over the 10 tasks, 5 seeds each, split by prior-history access (\textit{Task only} vs.\ \textit{Task + history}) and tool availability (with vs.\ w/o tools). \emph{ref.} denotes the fraction of calls resulting in a refusal and excluded from the analysis. Reasoning effort is \texttt{max} unless otherwise specified.}
  \label{tab:if-pass-rate}
  \centering
  \footnotesize
  \setlength{\tabcolsep}{4pt}
  \resizebox{\textwidth}{!}{%
  \begin{tabular}{l ccc ccc ccc ccc}
    \toprule
    & \multicolumn{6}{c}{\emph{Task only}} & \multicolumn{6}{c}{\emph{Task + history}} \\
    \cmidrule(lr){2-7} \cmidrule(lr){8-13}
    & \multicolumn{3}{c}{w/o tools} & \multicolumn{3}{c}{with tools}
    & \multicolumn{3}{c}{w/o tools} & \multicolumn{3}{c}{with tools} \\
    \cmidrule(lr){2-4} \cmidrule(lr){5-7} \cmidrule(lr){8-10} \cmidrule(lr){11-13}
    Model (effort)
      & ref. (\%) & strict & partial & ref. (\%) & strict & partial
      & ref. (\%) & strict & partial & ref. (\%) & strict & partial \\
    \midrule
    Claude Opus 5          & 12 & 0.87 & 0.96 & 28 & \textbf{1.00} & 1.00 & 10 & 0.83 & 0.95 & 52 & 0.98 & 0.99 \\
    Claude Opus 5 (xhigh)  &  0 & 0.83 & 0.95 &  2 & \textbf{1.00} & 1.00 &  2 & 0.67 & 0.88 & 12 & \textbf{1.00} & 1.00 \\
    Claude Opus 4.8        & 40 & 0.82 & 0.95 & 42 & 0.98 & 0.99 & 38 & 0.51 & 0.83 & 60 & \textbf{1.00} & 1.00 \\
    GPT-5.6-sol            &  0 & 0.74 & 0.90 &  0 & \textbf{1.00} & 1.00 &  0 & 0.54 & 0.82 &  0 & \textbf{1.00} & 1.00 \\
    GPT-5.6-terra          &  0 & 0.73 & 0.90 &  0 & \textbf{1.00} & 1.00 &  0 & 0.50 & 0.81 &  0 & \textbf{1.00} & 1.00 \\
    GPT-5.6-luna           &  0 & 0.73 & 0.89 &  0 & 0.99 & 1.00 &  0 & 0.60 & 0.84 &  0 & \textbf{1.00} & 1.00 \\
    \bottomrule
  \end{tabular}}
\end{table*}

\begin{figure}
  \centering
  \includegraphics[width=0.62\textwidth]{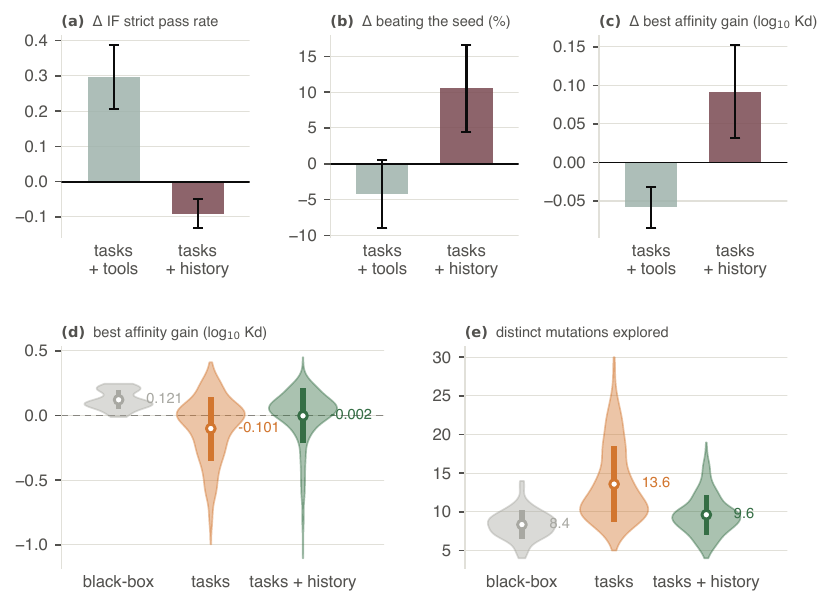}
\caption{\textbf{Tools improve compliance, history improves oracle performance, and
tasks explore where the black-box exploits.} \textbf{(a--c)} Main effects
of tools and of history, both measured for a task agent. 
Bars are the mean across six LLM models, whiskers $\pm1$ SD across models. \textbf{(d--e)} Distributions across task-seed batches for the three conditions, pooled over both tool conditions. Each batch contains 10 variants; violins show the mean and $\pm1$ SD, with means labelled. \textbf{(d)} Largest affinity gain per batch. \textbf{(e)} Number of distinct substitutions from the seed per batch.}\label{fig:IFandOracle}
\end{figure}

\paragraph{Tools and history provide complementary benefits, as do task-based and black-box settings.} In the task setting, we evaluate all four combinations of tool access and scored design history to isolate the contribution of each (see Figure~\ref{fig:IFandOracle}a--c). 
The two factors move in opposite directions: tools raise strict compliance by $+0.3$ while lowering both oracle measures, by $4.2$ points on candidates beating the seed and $0.06$ on best affinity gain, and history does the reverse, $-0.09$ on compliance, $+10.5$ points on beating the seed, and $+0.09$ on best affinity gain.

Figure~\ref{fig:IFandOracle}d--e shows that the black-box setting consistently improves over its seed, whereas task-guided edits reduce oracle score on average. However, task  edits have a longer positive tail, yielding larger gains when they succeed. Adding history raises the task setting mean to near zero. Task-guided trajectories also explore more substitutions on average ($13.6$) than black-box trajectories ($8.4$), with history intermediate ($9.6$).
Therefore, we conclude that the black-box setting favors exploitation, while task briefs encourage exploration and we used both in the design loop.

 \paragraph{Tools do not help in the black-box setting.} We run the black-box setting with and without tools on the same 5 seeds (Figure~\ref{fig:black-box task}). Tool access reduces the share of candidates that outperform the seed across all configurations, from 42--59\%  without tools to 22--38\% with tools. 
 The effect on best-of-batch gain is less consistent: all three GPT configurations decline, 
 whereas Claude Opus~5 and Opus~4.8 are flat or
slightly higher with tools and only Opus~5 (xhigh) declines. Since tool access does not consistently improve any metric, we omit tools from the black-box setting in the design loop.

\begin{figure}
  
  \centering
  \includegraphics[width=0.8\textwidth]{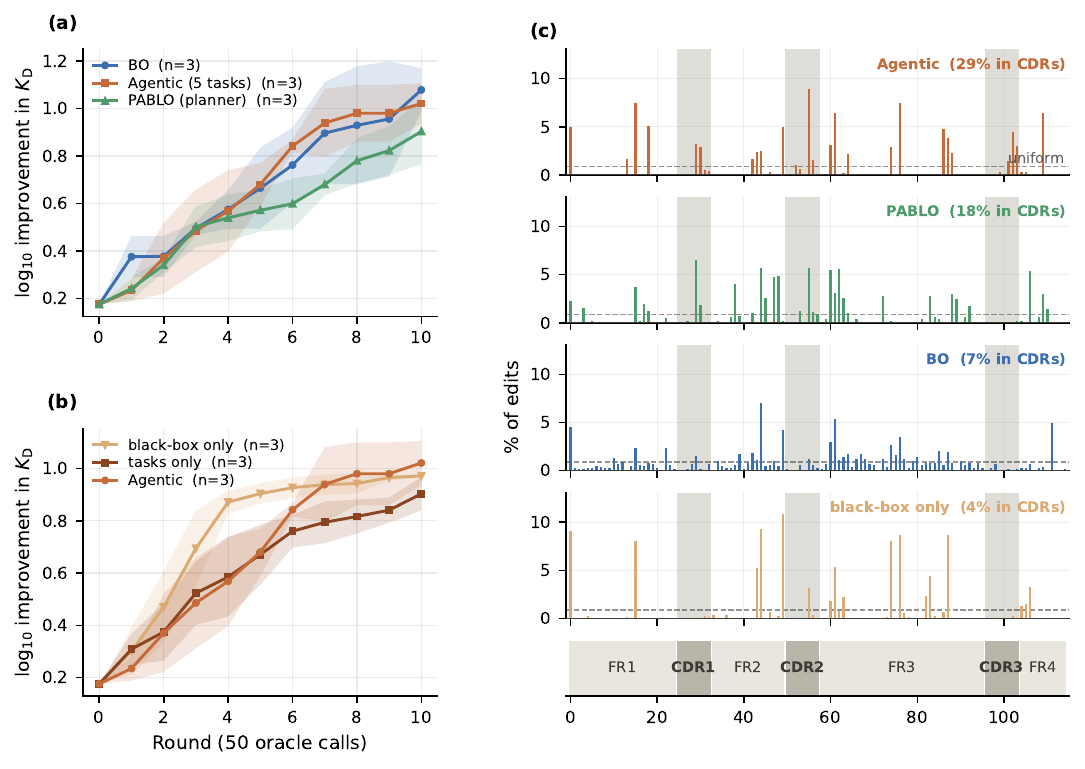}
\caption{\textbf{Three optimizers reach comparable affinity by editing different
parts of the domain.} All arms share a 50-sequence initialization pool and spend
10 rounds of 50 oracle calls on the same objective; lines are the mean over three
runs, bands $\pm1$ s.d. \textbf{(a)} Affinity trajectories for Bayesian
optimization over a GP surrogate, PABLO and the proposed agentic method. \textbf{(b)} Ablation of the proposed agentic method: \emph{black-box only} uses scored history, while \emph{tasks only} uses the four design briefs. \textbf{(c)} Distribution of edits in the final design round; the dashed line denotes the uniform baseline.}
\label{fig:drs}
\end{figure}

\paragraph{Comparable affinity, different edit distributions.}
We next run an in silico design loop from a simple system built out of our agents, as well as two baselines. We combined the black-box agent with agents using the top-four tasks and pooled their designs to decide what to characterize each round. PABLO \cite{maus2026pablo} and a Bayesian Optimization method \cite{ober2025sequence} are used as baselines. All three optimizers converge to within
$0.18$ decades of each other over 10 rounds, with overlapping bands throughout
(Figure~\ref{fig:drs}a). In the latest design round, the proposed agentic method edits CDRs 29\% of the time, reflecting the task instructions, compared with 18\% for PABLO and 7\% for Bayesian optimization  (Figure~\ref{fig:drs}c). Furthermore, ablating the black-box component reduces performance (Figure~\ref{fig:drs}b), whereas ablating the task component yields performance similar to the full agentic method. However, the resulting black-box-only edits are concentrated in a few regions (Figure~\ref{fig:drs}c), consistent with the exploit-heavy behavior.

\section{Discussion \& future work}
We have shown that a simple agentic optimizer can be effective for antibody design and examined the factors contributing to its performance. We curated a set of drug-design relevant tasks that are programmatically verifiable and had agents execute them. We showed that LLM judges agree strongly with programmatic verifiers. Design agents had inconsistent instruction following, but this was remedied with access to the proper tools. Surprisingly, making tools available reduced oracle performance. On the other hand, giving previously-scored variants had the opposite effect: it lowered instruction-following rates but improved the performance of generated candidates. We also evaluated a black-box setting in which the agent received only prior scored sequences. These agents were more exploit than explore-oriented and did not benefit from tools. We combined these agents into a multiagent system and ran simulated design rounds against an established Bayesian optimization baseline, finding comparable performance between the two.

Our study is currently limited to in-silico evaluation and a single affinity objective. Future work will extend to lab-validation and accommodate multi-objective design. Many avenues exist for implementing these extensions and we believe this work highlights the opportunity for precise molecular design agents that can cooperate effectively with scientists and engineers.

\newpage
\small
\bibliographystyle{plainnat}
\bibliography{references}

@inproceedings{zheng2025automation,
  title={From Automation to Autonomy: A Survey on Large Language Models in Scientific Discovery},
  author={Zheng, Tianshi and Deng, Zheye and Tsang, Hong Ting and Wang, Weiqi and Bai, Jiaxin and Wang, Zihao and Song, Yangqiu},
  booktitle={Proceedings of the 2025 Conference on Empirical Methods in Natural Language Processing},
  pages={17733--17750},
  year={2025},
  address={Suzhou, China},
  publisher={Association for Computational Linguistics}}

@article{lu2024aiscientist,
  title={The {AI} Scientist: Towards Fully Automated Open-Ended Scientific Discovery},
  author={Lu, Chris and Lu, Cong and Lange, Robert Tjarko and Foerster, Jakob and Clune, Jeff and Ha, David},
  journal={arXiv preprint arXiv:2408.06292},
  year={2024}
}

@inproceedings{
panigrahi2026heurekabench,
title={HeurekaBench: A Benchmarking Framework for {AI} Co-scientist},
author={Siba Smarak Panigrahi and Jovana Videnovi{\'c} and Maria Brbic},
booktitle={The Fourteenth International Conference on Learning Representations},
year={2026},
url={https://openreview.net/forum?id=Y7xCdFuFw7}
}

@article{kim2026biodesignbench,
  title={Benchmarking and behavioral characterization of {LLM} agents for protein design},
  author={Kim, Jeonghyeon and Romero, Philip},
  journal={bioRxiv},
  year={2026},
  note={Preprint, doi:10.64898/2026.05.06.723381}
}

@inproceedings{
wang2025large,
title={Large Language Model is Secretly a Protein Sequence Optimizer},
author={Yinkai Wang and Jiaxing He and Yuanqi Du and Xiaohui Chen and Jianan Canal Li and Liping Liu and Xiaolin Xu and Soha Hassoun},
booktitle={Learning Meaningful Representations of Life (LMRL) Workshop at ICLR 2025},
year={2025}
}

@article{maus2026pablo,
  title={Purely Agentic Black-Box Optimization for Biological Design},
  author={Maus, Natalie and Zeng, Yimeng and Jones, Haydn Thomas and Huang, Yining and Goel, Gaurav Ng and Rose, Alden and Kim, Kyurae and Lee, Hyun-Su and Der Torossian Torres, Marcelo and Wan, Fangping and de la Fuente-Nunez, Cesar and Yatskar, Mark and Bastani, Osbert and Gardner, Jacob R},
  journal={arXiv preprint arXiv:2601.22382},
  year={2026}
}

@article{wang2025swarms,
  title={Swarms of Large Language Model Agents for Protein Sequence Design with Experimental Validation},
  author={Wang, Fiona Y and Lee, Di Sheng and Kaplan, David L and Buehler, Markus J},
  journal={arXiv preprint arXiv:2511.22311},
  year={2025}
}

@article{jacob2025geniecat,
  title={Beyond Protein Language Models: An Agentic {LLM} Framework for Mechanistic Enzyme Design},
  author={Jacob, Bruno and Agarwal, Khushbu and Baer, Marcel and Rice, Peter and Raugei, Simone},
  journal={arXiv preprint arXiv:2511.19423},
  year={2025}
}

@article{ghafarollahi2024protagents,
  title={{ProtAgents}: protein discovery via large language model multi-agent collaborations combining physics and machine learning},
  author={Ghafarollahi, Alireza and Buehler, Markus J},
  journal={Digital Discovery},
  volume={3},
  number={7},
  pages={1389--1409},
  year={2024},
  publisher={Royal Society of Chemistry},
  doi = {10.1039/d4dd00013g}
}

@inproceedings{ge2026protocycle,
  title={ProtoCycle: Reflective Tool-Augmented Planning for Text-Guided Protein Design},
  author={Ge, Yutang and Zhao, Guojiang and Li, Sihang and Cheng, Zheng and Zhao, Zifeng and Xia, Hanchen and Ke, Guolin and Zhang, Linfeng and Gao, Zhifeng and Wang, Yu Guang},
  booktitle={Findings of the Association for Computational Linguistics: ACL 2026},
  pages={15562--15586},
  year={2026},
  doi={10.18653/v1/2026.findings-acl.763}
}

@article{xu2026proteinmcp,
  title={ProteinMCP: An agentic AI framework for autonomous protein engineering},
  author={Xu, Xiaopeng and Feng, Chenjie and Zha, Chao and He, Wenjia and He, Maolin and Xiao, Bin and Gao, Xin},
  journal={Protein Science},
  volume={35},
  number={4},
  pages={e70547},
  year={2026},
  doi={10.1002/pro.70547}
}

@inproceedings{
teneggi2026agentrosetta,
title={Protein Design with Agent Rosetta: A Case Study for Specialized Scientific Agents},
author={Jacopo Teneggi and SM Bargeen Alam Turzo and Tanya Marwah and Alberto Bietti and P. Douglas Renfrew and Vikram Khipple Mulligan and Siavash Golkar},
booktitle={Forty-third International Conference on Machine Learning},
year={2026}
}

@inproceedings{
qi2025agentif,
title={{AGENTIF}: Benchmarking Large Language Models Instruction Following Ability in Agentic Scenarios},
author={Yunjia Qi and Hao Peng and Xiaozhi Wang and Amy Xin and Youfeng Liu and Bin Xu and Lei Hou and Juanzi Li},
booktitle={The Thirty-ninth Annual Conference on Neural Information Processing Systems Datasets and Benchmarks Track},
year={2025},
url={https://openreview.net/forum?id=FLiMxTkIeu}
}

@article{su2026sciif,
  title={{SciIF}: Benchmarking Scientific Instruction Following Towards Rigorous Scientific Intelligence},
author={Encheng Su and Jianyu Wu and Chen Tang and Lintao Wang and Pengze Li and Aoran Wang and Jinouwen Zhang and Yizhou Wang and Yuan Meng and Xinzhu Ma and Shixiang Tang and Houqiang Li},
  journal={arXiv preprint arXiv:2601.04770},
  year={2026}
}

@article{he2026ori,
  title={Functional Protein Design and Enhancement with Ontology Reinforcement Iteration},
  author={He, Bing and Qin, Chenchen and Zhao, Yu and Huang, Long-Kai and Wu, Zihan and Wang, Fang and Wu, Fandi and Yang, Fan and Yao, Jianhua},
  journal={Nature Communications},
  volume={17},
  number={1},
  pages={4158},
  year={2026},
  doi={10.1038/s41467-026-69855-6}
}

@inproceedings{zhu2026maprot,
  title={Advancing Protein Design via Multi-Agent Reinforcement Learning with Pareto-Based Collaborative Optimization},
  author={Zhu, Mingming and Rao, Jiahua and Chen, Xiaoyu and Yuan, Qianmu and Yang, Yuedong},
  booktitle={Proceedings of the AAAI Conference on Artificial Intelligence},
  volume={40},
  number={2},
  pages={1650--1658},
  year={2026},
  doi={10.1609/aaai.v40i2.37142}
}

@inproceedings{shen2024toursynbio,
  title={TourSynbio: A Multi-Modal Large Model and Agent Framework to Bridge Text and Protein Sequences for Protein Engineering},
  author={Shen, Yiqing and Chen, Zan and Mamalakis, Michail and Liu, Yungeng and Li, Tianbin and Su, Yanzhou and He, Junjun and Li{\`o}, Pietro and Wang, Yu Guang},
  booktitle={2024 IEEE International Conference on Bioinformatics and Biomedicine (BIBM)},
  pages={2382--2389},
  year={2024},
  publisher={IEEE},
  doi={10.1109/BIBM62325.2024.10822695}
}

@inproceedings{
chen2025scienceagentbench,
title={ScienceAgentBench: Toward Rigorous Assessment of Language Agents for Data-Driven Scientific Discovery},
author={Ziru Chen and Shijie Chen and Yuting Ning and Qianheng Zhang and Boshi Wang and Botao Yu and Yifei Li and Zeyi Liao and Chen Wei and Zitong Lu and Vishal Dey and Mingyi Xue and Frazier N. Baker and Benjamin Burns and Daniel Adu-Ampratwum and Xuhui Huang and Xia Ning and Song Gao and Yu Su and Huan Sun},
booktitle={The Thirteenth International Conference on Learning Representations},
year={2025},
url={https://openreview.net/forum?id=6z4YKr0GK6}
}

@inproceedings{
fa2026bioagent,
title={BioAgent Bench: An {AI} Agent Evaluation Suite for Bioinformatics},
author={Dionizije Fa and Marko {\v{C}}uljak and Bruno Pand{\v{z}}a and Mateo {\v{C}}upi{\'c}},
booktitle={Forty-third International Conference on Machine Learning},
year={2026},
url={https://openreview.net/forum?id=2iFauBXg7Y}
}

@inproceedings{
    jansen2024discoveryworld,
    title={DiscoveryWorld: A Virtual Environment for Developing and Evaluating Automated Scientific Discovery Agents},
    author={Peter Jansen and Marc-Alexandre C{\^o}t{\'e} and Tushar Khot and Erin Bransom and Bhavana Dalvi Mishra and Bodhisattwa Prasad Majumder and Oyvind Tafjord and Peter Clark},
    booktitle={The Thirty-eight Conference on Neural Information Processing Systems Datasets and Benchmarks Track},
    year={2024},
    url={https://openreview.net/forum?id=cDYqckEt6d}
}

@article{zhou2023ifeval,
  title={Instruction-Following Evaluation for Large Language Models},
  author={Zhou, Jeffrey and Lu, Tianjian and Mishra, Swaroop and Brahma, Siddhartha and Basu, Sujoy and Luan, Yi and Zhou, Denny and Hou, Le},
  journal={arXiv preprint arXiv:2311.07911},
  year={2023},
  doi={10.48550/arXiv.2311.07911}
}

@inproceedings{jiang2024followbench,
  title={FollowBench: A Multi-level Fine-grained Constraints Following Benchmark for Large Language Models},
  author={Jiang, Yuxin and Wang, Yufei and Zeng, Xingshan and Zhong, Wanjun and Li, Liangyou and Mi, Fei and Shang, Lifeng and Jiang, Xin and Liu, Qun and Wang, Wei},
  booktitle={Proceedings of the 62nd Annual Meeting of the Association for Computational Linguistics (Volume 1: Long Papers)},
  pages={4667--4688},
  year={2024},
  doi={10.18653/v1/2024.acl-long.257}
}

@inproceedings{zhang2025cfbench,
  title={{CFB}ench: A Comprehensive Constraints-Following Benchmark for {LLM}s},
  author={Zhang, Tao and Zhu, Chenglin and Shen, Yanjun and Luo, Wenjing and Zhang, Yan and Liang, Hao and Zhang, Tao and Yang, Fan and Lin, Mingan and Qiao, Yujing and Chen, Weipeng and Cui, Bin and Zhang, Wentao and Zhou, Zenan},
  booktitle={Proceedings of the 63rd Annual Meeting of the Association for Computational Linguistics (Volume 1: Long Papers)},
  pages={32926--32944},
  year={2025},
  publisher={Association for Computational Linguistics},
  doi={10.18653/v1/2025.acl-long.1581}
}

@article{dunbar2016anarci,
  title={ANARCI: antigen receptor numbering and receptor classification},
  author={Dunbar, James and Deane, Charlotte M.},
  journal={Bioinformatics},
  volume={32},
  number={2},
  pages={298--300},
  year={2016},
  publisher={Oxford University Press},
  doi={10.1093/bioinformatics/btv552}
}

@article{abanades2023immunebuilder,
  title={ImmuneBuilder: Deep-Learning models for predicting the structures of immune proteins},
  author={Abanades, Brennan and Wong, Wing Ki and Boyles, Fergus and Georges, Guy and Bujotzek, Alexander and Deane, Charlotte M.},
  journal={Communications Biology},
  volume={6},
  number={1},
  pages={575},
  year={2023},
  publisher={Nature Publishing Group},
  doi={10.1038/s42003-023-04927-7}
}

@article{mitternacht2016freesasa,
  title={FreeSASA: An open source C library for solvent accessible surface area calculations},
  author={Mitternacht, Simon},
  journal={F1000Research},
  volume={5},
  pages={189},
  year={2016},
  doi={10.12688/f1000research.7931.1}
}

@article{hadsund2024nanobert,
  title={nanoBERT: a deep learning model for gene agnostic navigation of the nanobody mutational space},
  author={Hadsund, Johannes Thorling and Satława, Tadeusz and Janusz, Bartosz and Shan, Lu and Zhou, Li and Röttger, Richard and Krawczyk, Konrad},
  journal={Bioinformatics Advances},
  volume={4},
  number={1},
  pages={vbae033},
  year={2024},
  publisher={Oxford University Press},
  doi={10.1093/bioadv/vbae033}
}

@article{cock2009biopython,
  title={Biopython: freely available Python tools for computational molecular biology and bioinformatics},
  author={Cock, Peter J. A. and Antao, Tiago and Chang, Jeffrey T. and Chapman, Brad A. and Cox, Cymon J. and Dalke, Andrew and Friedberg, Iddo and Hamelryck, Thomas and Kauff, Frank and Wilczynski, Bartek and de Hoon, Michiel J. L.},
  journal={Bioinformatics},
  volume={25},
  number={11},
  pages={1422--1423},
  year={2009},
  publisher={Oxford University Press},
  doi={10.1093/bioinformatics/btp163}
}

@inproceedings{
ober2025sequence,
title={Is Sequence Information All You Need for Bayesian Optimization of Antibodies?},
author={Sebastian W. Ober and Calvin McCarter and Aniruddh Raghu and Yucen Lily Li and Alan Nawzad Amin and Andrew Gordon Wilson and Hunter Elliott},
booktitle={NeurIPS 2025 AI for Science Workshop},
year={2025},
url={https://openreview.net/forum?id=wYuUqaSazT}
}


\appendix
\section{Related Work}

\textbf{Agents for protein design.} Agents can coordinate multi-step protein design workflows by calling tools across the standard protein design stack. \citep{ghafarollahi2024protagents, kim2026biodesignbench, xu2026proteinmcp, teneggi2026agentrosetta}. 
Recent multi-agent systems extend this paradigm beyond simple tool use: ProtoCycle combines iterative planning, execution, and reflection with tool feedback \citep{ge2026protocycle}, while ORI combines generation, evaluation, and experimental feedback in a closed-loop workflow \citep{he2026ori}. Multi-agent frameworks such as ProtAgents, TourSynbio, and MAProt use specialized agents as planners, optimizers, and coordinators of autonomous design workflows \citep{ghafarollahi2024protagents, shen2024toursynbio, zhu2026maprot}.

\textbf{Instruction following and constraint compliance.} A broad line of work evaluates whether LLMs adhere to explicit instructions. IFEval introduced verifiable atomic constraints \citep{zhou2023ifeval}; FollowBench and CFBench extended this to multi-constraint composition \citep{jiang2024followbench, zhang2025cfbench}. 
Recent benchmarks evaluate task success and instruction following separately. In agentic settings, AGENTIF shows that models struggle to satisfy complex constraints and tool specifications \citep{qi2025agentif}, while scientific-agent benchmarks reveal persistent failures in reliably executing multi step scientific workflows \citep{jansen2024discoveryworld, chen2025scienceagentbench,fa2026bioagent}. SciIF brings this lens directly to scientific instruction following, scoring answer correctness and constraint compliance separately and showing that models can be “right for the wrong reasons,” producing correct outcomes while failing to satisfy the evidence or constraints required for scientific validity \citep{su2026sciif}.

\section{Design tasks}\label{appendix:design-tasks}

\begin{taskcard}{CHARGE REPAIR}
\cardtop{Summary}
When a seed has an isoelectric point that is too low or high, towards the desired soluble range of 7.5-9.0 by editing surface charge; when the seed is already safe, diversify normally.\\

\cardtop{Must-haves}
\begin{mh}
  \item The output's total sequence length is exactly equal to the seed's total sequence length.
  \item The two conserved framework cysteines and the four FR2 hallmark positions (Kabat 37, 44, 45, 47) are unchanged from the seed.
  \item On the predicted structure, every position whose charge changed from the seed (a cationic K/R or anionic D/E residue gained or lost) is solvent-exposed (relative SASA at least 0.25), and no buried framework position was altered.
  \item Conditional action gate: if the seed's isoelectric point is outside [7.5, 9.0], then at least 1 substitution brings the output's isoelectric point inside [7.5, 9.0]. If the seed's isoelectric point is already inside [7.5, 9.0], then at least 1 residue differ from the seed and the output's isoelectric point is still inside [7.5, 9.0].
\end{mh}
\cardtop{Suggestions (optional)}
\begin{sg}
  \item Neutralize on the framework surface (swap an exposed K/R or D/E for Q/N/S/T), leaving the paratope for the directional tasks -- or trim one charged CDR residue instead. Split the batch across the two.
  \item Vary the dose: some variants one decisive swap, others several small ones.
\end{sg}
\end{taskcard}

\begin{taskcard}{PARATOPE ELECTROSTATICS}
\cardtop{Summary}
Put a coherent, single-sign charged patch on the CDR surface so the paratope can complement a charged epitope, a primary driver of affinity and specificity. The task does not pick a sign -- a batch presents positive and negative paratopes and lets the oracle choose. If a PRIOR SCORED VARIANTS block appears below, use it to infer which sign the oracle has favored.\\

\cardtop{Must-haves}
\begin{mh}
  \item The output's total sequence length is exactly equal to the seed's total sequence length.
  \item The pooled CDRs contain at least 3 charged residues (cationic = K/R, anionic = D/E).
  \item The charged CDR residues are predominantly one sign: the majority sign is at least two-thirds of all charged CDR residues.
  \item At least 1 residue in the pooled CDRs differ from the seed.
\end{mh}
\cardtop{Suggestions (optional)}
\begin{sg}
  \item Minimal patch (aromatics/H-bonders carry the contacts) in some variants; a bold, dense patch in others -- let the oracle judge which the epitope wants.
  \item Vary sign, amount, and one-loop-vs-spread across variants. Don't assume positive/heavy/clustered is best.
  \item Place each charged residue next to a contact-maker (Y, S, or N).
\end{sg}
\end{taskcard}

\begin{taskcard}{BACKBONE SHAPE}
\cardtop{Summary}
Reshape a CDR loop by adding or removing a proline or a glycine -- the two residues that most control a loop's backbone. A proline stiffens a loop into a defined kink; a glycine loosens it into a flexible hinge.\\

\cardtop{Must-haves}
\begin{mh}
  \item The output's total sequence length is exactly equal to the seed's total sequence length.
  \item At least 1 residue in the pooled CDRs differ from the seed.
  \item The prolines and glycines in the CDR loops sit differently than in the seed: at least one CDR position changes between proline, glycine, and other in a way that adds, removes, or swaps a proline or glycine (a glycine-to-proline swap counts).
  \item Every proline/glycine change stays inside a CDR loop; the framework's own prolines and glycines are left exactly as the seed has them.
\end{mh}
\cardtop{Suggestions (optional)}
\begin{sg}
  \item Add a proline to stiffen a loop into a fixed shape -- or a glycine to loosen it into a flexible hinge. Split the batch across both.
  \item If the seed loop already holds a P/G, remove it and let the loop relax open.
  \item Placement matters: a P/G at the loop base reshapes how the loop is presented; near the tip, it reshapes the contact face.
\end{sg}
\end{taskcard}

\begin{taskcard}{SURFACE TARGETED}
\cardtop{Summary}
Focus binding changes on the solvent-exposed CDR positions that contact antigen, and leave the buried core that holds the fold together untouched.\\

\cardtop{Must-haves}
\begin{mh}
  \item The output's total sequence length is exactly equal to the seed's total sequence length.
  \item At least 1 residue in the pooled CDRs differ from the seed.
  \item On the predicted structure, every position that differs from the seed has relative solvent accessibility at least 25\% -- only exposed positions were changed; no buried-core position was mutated.
\end{mh}
\cardtop{Suggestions (optional)}
\begin{sg}
  \item Spread edits across many exposed positions for a broad interface -- or concentrate them into one focused exposed hot-patch. Split the batch across both.
  \item At exposed positions, prefer polar/charged/aromatic over burying a new hydrophobic.
\end{sg}
\end{taskcard}
\begin{taskcard}{PARATOPE COMPOSITION}
\cardtop{Summary}
Bias the three CDRs toward the residue types that dominate natural antibody binding sites -- a tyrosine-and-serine-led surface that covers antigen well while staying developable.\\

\cardtop{Must-haves}
\begin{mh}
  \item The output's total sequence length is exactly equal to the seed's total sequence length.
  \item Across the pooled CDRs, tyrosine and serine together are at least 25\% of all CDR residues.
  \item No single residue type is more than 40\% of the pooled CDR residues.
  \item At least 1 residue in the pooled CDRs differ from the seed.
\end{mh}
\cardtop{Suggestions (optional)}
\begin{sg}
  \item Aromatic-heavy (Y/W/F) for a flat, sticky surface -- or aromatic-light and small-polar for a smaller, flexible one. Split the batch across both.
  \item In some variants add fresh aromatics; in others keep the seed's aromatics and edit only the plain residues around them.
\end{sg}
\end{taskcard}

\begin{taskcard}{H BOND NETWORK}
\cardtop{Summary}
Build a paratope that recognizes its target mainly through hydrogen bonds rather than charge or greasy packing -- the route to a specific, discriminating binder. Enrich the loops in the polar residues that donate and accept hydrogen bonds, and keep them varied so the network is real rather than a monotonous run of one residue.\\

\cardtop{Must-haves}
\begin{mh}
  \item The output's total sequence length is exactly equal to the seed's total sequence length.
  \item At least 1 residue in the pooled CDRs differ from the seed.
  \item Across the pooled CDRs, the hydrogen-bonding polar residues (serine, threonine, asparagine, glutamine -- S/T/N/Q) together make up at least 30\% of all CDR residues.
  \item No single residue type is more than 40\% of the pooled CDR residues, so the loops don't collapse into a poly-serine stretch.
\end{mh}
\cardtop{Suggestions (optional)}
\begin{sg}
  \item Cluster the polar residues on one loop for a focused contact -- or spread them across all three CDRs for a broad polar surface. Try both.
  \item N/Q for strong anchors (two H-bonds each); S/T for a lighter, adjustable network.
  \item Pair a polar residue with an adjacent aromatic (Y/W) -- packing plus a locked H-bond. Tyrosine can do both.
  \item Keep new N out of Asn-Gly/Asn-Ser and Asn-X-Ser/Thr.
\end{sg}
\end{taskcard}

\begin{taskcard}{AGGREGATION CONTROL}
\cardtop{Summary}
Keep the CDR loops free of hydrophobic stretches, which drive aggregation and sticky nonspecific binding. Actively substitute -- do not return the seed unchanged.\\

\cardtop{Must-haves}
\begin{mh}
  \item The output's total sequence length is exactly equal to the seed's total sequence length.
  \item At least 1 residue in the pooled CDRs differ from the seed.
  \item No run of more than 3 hydrophobic residues in a row in any CDR.
  \item The hydrophobic fraction of CDR3 is at most 0.45.
\end{mh}
\cardtop{Suggestions (optional)}
\begin{sg}
  \item Break a hydrophobic run with a polar or charged residue.
  \item Keep exposed hydrophobics off the loop tips (buried ones are fine).
\end{sg}
\end{taskcard}

\begin{taskcard}{CDR DIVERSIFY BASELINE}
\cardtop{Summary}
Diversify the CDR loops with no particular direction but in line with natural variation -- no charge push, no composition target, no favored loop -- just make a handful of edits spread across the paratope.\\

\cardtop{Must-haves}
\begin{mh}
  \item The output's total sequence length is exactly equal to the seed's total sequence length.
  \item Every CDR keeps the seed's length -- substitutions only, no residues inserted or deleted in any loop.
  \item At least 1 residue in the pooled CDRs differ from the seed; if 2 or more differ, they are not all in one loop -- they are spread across at least two of the three CDRs.
\end{mh}
\cardtop{Suggestions (optional)}
\begin{sg}
  \item Vary the number of edits per variant (light to heavy) so the batch spans a range of distances from the seed.
  \item Let variants drift different ways by chance -- none required to lean anywhere.
  \item Touch CDR1/2 as readily as CDR3; don't collapse onto CDR3 by habit.
\end{sg}
\end{taskcard}

\begin{taskcard}{MUTATION RECOMBINATION}

\cardtop{Summary}
Given the seed and the PRIOR SCORED VARIANTS block below -- variants oracle-evaluated in earlier iterations of this run, treated as reference variants in the rules below (higher oracle score = higher affinity) -- combine substitutions from those references into combinations present in no single reference. Exploit operator: invents no new mutations; recombines already-tolerated ones. How many to combine is open -- small subsets and large stacks both valid, because good mutations often combine non-additively.\\

\cardtop{Must-haves}
\begin{mh}
  \item Each output is the same length as the seed.
  \item Every position where the output differs from the parental matches a substitution (same position, same residue) carried by one of the reference variants shown.
  \item Each output carries at least 2 such substitutions, and no single reference variant carries all of them.
  \item Each output is not identical to the seed, the parental, or any reference variant.
\end{mh}
\end{taskcard}

\begin{taskcard}{RIGIDIFYING DISULFIDE}
\cardtop{Summary}
Rigidify a flexible CDR loop -- especially a long CDR3 -- by adding a matched cysteine pair that forms a disulfide staple, pre-organizing the paratope so it pays less entropy on binding.\\

\cardtop{Must-haves}
\begin{mh}
  \item The output's total sequence length is exactly equal to the seed's total sequence length.
  \item At least 1 residue in the pooled CDRs differ from the seed.
  \item The output introduces exactly one new pair of cysteines (two added) within the CDRs relative to the seed, beyond the two conserved framework cysteines.
  \item On the predicted structure, the two introduced cysteines form a disulfide (S-gamma to S-gamma distance within 1.8-2.5 Angstrom), leaving no free thiol.
\end{mh}
\cardtop{Suggestions (optional)}
\begin{sg}
  \item Staple across the base of a long CDR3 to clamp one loop -- or bridge two loops (e.g.\ CDR1-CDR3) to lock their relative geometry. Split the batch across both.
  \item Vary the span: a short tight clamp in some variants, a wider bridge in others.
\end{sg}
\end{taskcard}

\begin{taskcard}{BLACK-BOX}
\textit{No design brief -- no SUMMARY, no MUST-HAVES, no SUGGESTIONS. The worker is shown the PRIOR SCORED VARIANTS block (prior sequences with their oracle scores, higher = better) plus the starting sequence and its own score, and must simply propose sequences that score higher.}

\cardtop{Rules (verbatim)}
\begin{mh}
  \item Study the shown sequences and their scores; every output MUST aim to score HIGHER than any of them.
  \item Every output MUST differ from the input sequence -- never return the input unchanged.
  \item PRESERVE THE TOTAL SEQUENCE LENGTH of the input on every output.
  \item Output 10 antibody variants per call.
  \item Make the variants DIVERSE: have them explore DIFFERENT modifications (including opposite directions) so the batch spans the design space rather than clustering on one solution.
  \item Base each design on the starting sequence, editing it toward the intent.
\end{mh}
\end{taskcard}

\section{Prompts}\label{appendix:prompts}

\begin{promptbox}{Task setting ( no tools, no history)}
=== SYSTEM PROMPT ===

You are an expert antibody designer (single-domain / nanobody).
You will receive an antibody design brief with three sections:
  SUMMARY     -- one-line statement of the design intent for this call.
  MUST-HAVES  -- hard requirements EVERY variant you output must satisfy.
  SUGGESTIONS -- optional design directions; they may point in mutually
                 exclusive directions, which is intentional.
INPUT: you will also be given a single starting antibody sequence.

Rules:
  - Every output MUST satisfy ALL must-haves. If a suggestion conflicts
    with a must-have, the must-have always wins.
  - Every output MUST differ from the input sequence -- never return the input unchanged.
  - PRESERVE THE TOTAL SEQUENCE LENGTH of the input on every output.
  - Every output differing from the input at more than 3 positions is
    DISCARDED before scoring and counts for nothing, even if it satisfies every
    must-have. Count your substitutions: where a must-have and this limit
    conflict, stay within the limit.
  - Output 10 antibody variants per call.
  - Make the variants DIVERSE: have them follow DIFFERENT suggestions
    (including opposite ends of any paired suggestions) so the batch
    spans the design space rather than clustering on one solution.
  - Base each design on the starting sequence, editing it toward the intent.

SUMMARY: Bias the three CDRs toward the residue types that dominate natural
antibody binding sites -- a tyrosine-and-serine-led surface that covers antigen
well while staying developable.

MUST-HAVES:
1. The output's total sequence length is exactly equal to the seed's total sequence length.
2. Across the pooled CDRs, tyrosine and serine together are at least 25% of all CDR residues.
3. No single residue type is more than 40% of the pooled CDR residues.
4. At least 1 residue in the pooled CDRs differ from the seed.

SUGGESTIONS (optional):
- Aromatic-heavy (Y/W/F) for a flat, sticky surface -- or aromatic-light and
  small-polar for a smaller, flexible one. Split the batch across both.
- In some variants add fresh aromatics; in others keep the seed's aromatics and
  edit only the plain residues around them.
- Consider checking whether your proposed mutations look plausible in nanobody
  sequence context.

OUTPUT FORMAT (REQUIRED): Return ONLY a JSON object with a list of 10 protein
sequences called 'candidates'.

=== USER TURN ===

Input Protein:
Modify it to generate 10 new protein sequences.
\end{promptbox}

\begin{promptbox}{+ tools}
## VERIFICATION TOOLS
You have deterministic inspection tools to CHECK your candidates before you
finalize them -- do not rely on eyeballing.
- Each tool MEASURES (counts, regions, charge, predicted structure); it does not
  tell you whether a candidate is acceptable. You read the numbers and decide.
- Batch your candidates: every tool's `sequences` argument is a list -- pass all
  your candidates in one call rather than one call per candidate.
- Each tool's own description says exactly what it returns and when to use it.
- The tools are for your own verification only. Your final answer must still be
  the variants in the required output format -- unchanged.
\end{promptbox}

\begin{promptbox}{+ history}
MUST-HAVES:
5. Given the PRIOR SCORED VARIANTS block below, each output is a deliberate
   attempt to score higher than the starting sequence: infer from the scored
   variants which changes the oracle rewards and apply that reasoning, rather
   than editing arbitrarily.

PRIOR SCORED VARIANTS:

Higher score = higher affinity.  Your starting sequence (Input Protein) scored -0.4210.

  parental:      (ancestral scaffold)

   1      score = -0.4210
   2      score = -0.3115

These variants were oracle-evaluated earlier in this run; do not re-propose
them. Use them -- with your tools if useful -- to infer what direction of change
correlates with better scores. Consider checking whether your proposed mutations
look plausible in nanobody sequence context.

=== USER TURN ===

Input Protein:    (oracle score: -0.4210)
\end{promptbox}

\begin{promptbox}{LLM Judge}
You are a strict binary judge scoring antibody (nanobody) variants an AI worker produced against a locked design brief. For each (output, must-have) pair, decide whether the OUTPUT satisfies the must-have.

## DESIGN BRIEF
SUMMARY: {summary}
MUST-HAVES:
{must_haves_block}

## SEED  (the antibody the worker was told to modify -- provenance for seed-relative must-haves; judge the OUTPUT.)
{seed}
## OUTPUTS  (score every one against every must-have; 1-indexed)
{outputs_block}

## RUBRIC (one verdict per (output, must-have) pair)
  PASS -- output satisfies the must-have.
  FAIL -- output violates the must-have.

## SCORING CONVENTIONS
- Residue groups: {residue_groups}. "Pooled CDRs" = CDR1+CDR2+CDR3.
- Ground every verdict in a tool call -- do not count residues, infer regions, or compare to the seed by eye. Each tool's description says what it is for.
- Some must-haves reference the predicted 3D structure -- solvent exposure of positions that changed from the seed, or whether cysteines are positioned to form a disulfide. Use fold_and_measure for these: it returns per-residue relative solvent accessibility (RSA, by IMGT label) and cysteine SG-SG distance. Apply the exact threshold the must-have states (e.g. its RSA cutoff for "exposed", or its SG-SG) distance range for a disulfide).
- Batch every sequence into ONE tool call -- do not call once per output.

## OUTPUT
Call submit_verdicts exactly ONCE with the full verdict grid (every output x every must-have). Return a verdict for EVERY pair; there are no gaps.

Each entry MUST include:
  - `must_have_index`: the 1-indexed must-have this verdict is about.
  - `must_have_text`:  the verbatim text of MUST-HAVE #<must_have_index> from the DESIGN BRIEF above. Copy the exact rule you are grading against so you don't apply an adjacent must-have's threshold.
  - `verdict`: PASS or FAIL, decided AGAINST the must_have_text you just copied.
  - `evidence`: numeric grounding from a tool call. Your evidence and verdict must be internally consistent -- if the evidence text says PASS then verdict must be PASS, and vice versa.

\end{promptbox}

\begin{promptbox}{Black-box setting}
You are an expert antibody designer (single-domain / nanobody).
You will receive:
  PRIOR SCORED VARIANTS -- variants oracle-evaluated in earlier iterations of
                           this run, with numerical objective scores
                           (higher = better).
INPUT: you will also be given a single starting antibody sequence.

Rules:
  - Study the shown sequences and their scores; every output MUST aim to
    score HIGHER than any of them.
  - Every output MUST differ from the input sequence -- never return the input unchanged.
  - PRESERVE THE TOTAL SEQUENCE LENGTH of the input on every output.
  - Every output differing from the input at more than 3 positions is
    DISCARDED before scoring and counts for nothing, even if it satisfies every
    must-have. Count your substitutions: where a must-have and this limit
    conflict, stay within the limit.
  - Output 10 antibody variants per call.
  - Make the variants DIVERSE: have them explore DIFFERENT modifications
    (including opposite directions) so the batch spans the design space
    rather than clustering on one solution.
  - Base each design on the starting sequence, editing it toward the intent.
\end{promptbox}

\section{Run configurations}\label{appendix:configs}

All three optimizers spend the same budget on the same objective: a 50-sequence
initialization pool shared across every run, then 10 rounds of 50 oracle calls.
Each agent call returns 10 candidate variants. Every candidate must preserve the
seed's total length, and candidates differing from their seed by more than the
arm's edit limit are discarded before scoring.

\paragraph{Bayesian optimization.} A Gaussian-process surrogate over the same
initialization pool and budget, run three times differing only in the
GA/acquisition random seed. Candidates more than 3 edits from the initialization
pool are filtered out.

\paragraph{Agentic.} A single worker, GPT-5.6~Sol at
\texttt{max} reasoning effort with adaptive thinking and a 128k token limit,
reading a fixed five-task brief: \texttt{CHARGE\_REPAIR},
\texttt{SURFACE\_TARGETED}, \texttt{PARATOPE\_ELECTROSTATICS},
\texttt{BACKBONE\_SHAPE}, and an unconstrained \texttt{BLACK-BOX} task. Candidates
may differ from their seed at up to 3 positions. Measurement tools are available,
the scored-history block is enabled, and scores are
taken as the minimum over the affinity ensemble. \texttt{BLACK-BOX} runs without
tools as the no-guidance control. In each round, two diverse seeds are selected greedily by score, with a minimum edit distance of 2. Agents then optimize each seed under the assigned task or black-box setting. For oracle evaluation, each (task, seed) arm contributes equally from its highest surrogate-scoring candidates. 

\paragraph{PABLO.} Three agents: an explorer and a planner, both GPT-5.6~Sol, and
a worker, GPT-5.6~Luna, all at \texttt{max} effort with adaptive thinking. The
planner selects each round's tasks rather than reading a fixed list, and the
explorer proposes globally with an edit radius of 1 and one retry. Each round starts from two diverse seeds and returns 50 oracle candidates, with the oracle budget divided equally across (seed, task) pairs. Cumulative edits per
round are capped at 3, matching the proposed agentic approach. One failed improvement attempt is allowed per trajectory: worker calls repeat until the surrogate score fails to improve once. Oracle scores are taken as the ensemble minimum. Measurement tools are disabled, consistent with the original paper. Prompts are minimally adapted for antibody design.

\section{Tool usage}\label{appendix:toolUsage}

\subsection{Tool usage metrics} \label{sec:toolUse}
Tool-use metrics are defined per worker call. The \emph{refusal rate} is the fraction
of attempted calls that returned no candidates; every other metric is averaged over
completed calls only, so a refused call counts toward refusals and nothing else.
\begin{itemize}
  \item \textbf{tool calls}: mean number of tool invocations per worker call.
  \item \textbf{measurements}: mean number of tool--sequence pairs per worker call,
        counted with multiplicity, so one sequence sent to three tools is three
        measurements.
  \item \textbf{re-measured}: fraction of measurements in which a tool was handed a
        sequence it had already been given. Since every tool is deterministic, this
        is wasted work.
  \item \textbf{invented}: mean number of distinct sequences per worker call that the
        worker generated itself, i.e.\ excluding the seed, the parental, and the
        provided references.
  \item \textbf{abandoned}: fraction of invented sequences that the worker measured
        but did not return. This is the cost of screening, not an error.
  \item \textbf{rounds}: mean number of propose--measure cycles per worker call,
        i.e.\ distinct turns in which the worker proposed new candidates and called
        tools on them.
\end{itemize}

\paragraph{Models use the tools heavily and productively.}
Every completed worker call issues 7 to 12 tool calls and 49 to 126 measurements
(Table~\ref{tab:tool-use}), so tool access is exercised rather than nominal, with low
redundancy (3 to 10\% re-measurement). Most remaining effort is screening rather than
error: models measure many more candidates than they return (37 to 63\% abandoned).
The models differ mainly in search style, with Claude Opus 5 reaching the same
compliance as the more exhaustive GPT-5.6-sol at far lower cost (48.7 versus 126.3
measurements), i.e.\ greater sample efficiency.

\begin{table*}[h]
  \caption{\textbf{Tool use across context conditions.} Tool-use metrics from
    Sec.~\ref{sec:toolUse}, mean $\pm$ one SD across completed worker calls. Refused
    calls are excluded from every metric column and appear only in \emph{ref}. Reasoning
    effort is max unless marked \texttt{xhigh}. The three arms differ in what the worker
    receives: \emph{task + history}, \emph{task only}, and \emph{black-box}.\\}
  \label{tab:tool-use}
  \centering
  \footnotesize
  \setlength{\tabcolsep}{3pt}
  \begin{tabular}{@{}l@{\hskip 6pt}r@{\hskip 6pt}rrrrrr@{}}
    \toprule
    & & \multicolumn{3}{c}{measurement effort} & \multicolumn{3}{c}{candidate search} \\
    \cmidrule(lr){3-5} \cmidrule(lr){6-8}
    Model & ref. (\%) & tool calls & measurements & re-meas.(\%)
                   & rounds & invented & discarded(\%) \\
    \midrule
    \multicolumn{8}{@{}l}{\emph{Task + history}} \\
    \midrule
    Claude Opus 5         & 52 & $6.3\pm1.3$  & $57.4\pm13.1$  & $8.6\pm10.0$  & $1.6\pm0.7$ & $14.9\pm6.7$  & $22.9\pm24.6$ \\
    Claude Opus 5 (xhigh) & 12 & $6.3\pm2.1$  & $57.2\pm17.3$  & $5.9\pm9.0$   & $2.1\pm0.7$ & $19.1\pm7.3$  & $41.6\pm23.4$ \\    
    Claude Opus 4.8       & 60 & $7.7\pm3.1$  & $79.0\pm23.1$  & $11.5\pm12.6$ & $1.9\pm0.9$ & $19.9\pm7.6$  & $43.4\pm22.7$ \\
    GPT-5.6-sol           & 0  & $10.8\pm4.2$ & $129.3\pm40.6$ & $8.4\pm7.1$   & $2.9\pm1.1$ & $39.9\pm14.2$ & $71.5\pm13.1$ \\
    GPT-5.6-terra         & 0  & $12.6\pm4.3$ & $133.7\pm50.4$ & $11.3\pm9.1$  & $3.1\pm1.1$ & $29.5\pm10.5$ & $60.0\pm20.0$ \\    
    GPT-5.6-luna          & 0  & $14.8\pm5.3$ & $142.4\pm39.8$ & $19.1\pm9.3$  & $5.2\pm2.3$ & $31.8\pm13.8$ & $62.5\pm16.0$ \\
    \midrule
    \multicolumn{8}{@{}l}{\emph{Task only}} \\
    \midrule
    Claude Opus 5         & 28 & $8.7\pm2.1$  & $63.4\pm20.8$  & $9.1\pm10.5$  & $2.5\pm0.8$ & $20.3\pm8.7$  & $43.0\pm22.7$ \\
    Claude Opus 5 (xhigh) & 2  & $6.8\pm2.0$  & $47.5\pm14.2$  & $7.2\pm9.0$   & $2.4\pm0.8$ & $18.3\pm6.0$  & $41.2\pm19.2$ \\
    Claude Opus 4.8       & 42 & $8.6\pm2.8$  & $64.2\pm17.5$  & $16.0\pm18.5$ & $2.3\pm0.7$ & $19.6\pm5.9$  & $44.7\pm17.4$ \\
    GPT-5.6-sol           & 0  & $10.8\pm4.7$ & $108.9\pm33.8$ & $9.8\pm8.2$   & $3.0\pm1.1$ & $35.7\pm12.1$ & $68.3\pm12.8$ \\
    GPT-5.6-terra         & 0  & $11.8\pm4.7$ & $94.7\pm37.1$  & $11.4\pm10.0$ & $2.5\pm1.0$ & $20.1\pm9.3$  & $40.1\pm25.0$ \\   
    GPT-5.6-luna          & 0  & $8.7\pm5.1$  & $69.6\pm47.3$  & $13.1\pm14.8$ & $2.3\pm1.7$ & $15.3\pm8.1$  & $21.4\pm26.7$ \\
    \midrule
    \multicolumn{8}{@{}l}{\emph{Black-box}} \\
    \midrule
    Claude Opus 5         & 20 & $6.2\pm0.4$  & $58.0\pm7.0$   & $0.8\pm0.9$   & $1.8\pm0.8$ & $14.0\pm6.4$  & $17.3\pm24.9$ \\
    Claude Opus 5 (xhigh) & 0  & $4.6\pm0.8$  & $48.6\pm8.7$   & $0.3\pm0.6$   & $1.2\pm0.4$ & $10.8\pm1.6$  & $7.1\pm14.3$  \\
    Claude Opus 4.8       & 0  & $11.2\pm1.9$ & $102.8\pm16.0$ & $6.5\pm5.5$   & $2.0\pm0.6$ & $14.4\pm2.0$  & $30.8\pm9.2$  \\
    GPT-5.6-sol           & 0  & $14.8\pm2.3$ & $174.6\pm13.3$ & $13.0\pm4.6$  & $4.8\pm1.0$ & $36.2\pm11.1$ & $71.1\pm8.2$  \\
    GPT-5.6-terra         & 0  & $13.8\pm2.6$ & $120.2\pm21.3$ & $14.0\pm8.5$  & $3.4\pm1.2$ & $23.8\pm5.3$  & $55.5\pm11.2$ \\    
    GPT-5.6-luna          & 0  & $15.4\pm4.3$ & $162.6\pm19.8$ & $16.1\pm10.5$ & $4.8\pm1.9$ & $22.8\pm10.1$ & $49.7\pm22.5$ \\
    \bottomrule
  \end{tabular}
\end{table*}

\begin{table}
  \caption{\textbf{Tool use on Rigidifying Disulfide.} The task excluded from
  Tab.~\ref{tab:tool-use}, reported on the same definitions. Its must-haves require
  both a correctly placed pair of new cysteines and a disulfide bond confirmed on the
  predicted structure, and it is the only task where instruction following without
  tools falls below $10\%$. Every model responds by generating far more candidates
  than on any other task -$3.8$ to $8.3$ times its own nine-task mean- and
  discarding $84$--$94\%$ of them.}
  \label{tab:disulfide}
  \centering
  \small
  \setlength{\tabcolsep}{4pt}
  \begin{tabular}{@{}lrrrrr@{}}
    \toprule
    Model (effort) & tool calls & re-meas.\ (\%) & rounds & invented & abandoned (\%) \\
    \midrule
    Claude Opus 5 (xhigh) & $10.0 \pm 1.8$ & $0.0 \pm 0.0$ & $5.0 \pm 1.3$ & $62.0 \pm 5.4$   & $83.7 \pm 1.4$ \\
    Claude Opus 5 (max)   & $10.6 \pm 1.2$ & $0.0 \pm 0.0$ & $5.0 \pm 1.1$ & $81.6 \pm 16.6$  & $87.2 \pm 2.6$ \\
    GPT-5.6-sol (max)     & $18.6 \pm 5.6$ & $6.7 \pm 2.0$ & $8.6 \pm 3.9$ & $238.4 \pm 136.1$ & $94.4 \pm 2.7$ \\
    GPT-5.6-terra (max)   & $18.4 \pm 4.2$ & $7.1 \pm 2.7$ & $7.8 \pm 1.5$ & $135.2 \pm 39.0$ & $91.9 \pm 2.5$ \\
    \bottomrule
  \end{tabular}
\end{table}
\section{Breaking down instruction following failures}\label{appendix:if-failure}
\begin{figure}[h]
  \centering
  \includegraphics[width=0.65\textwidth]{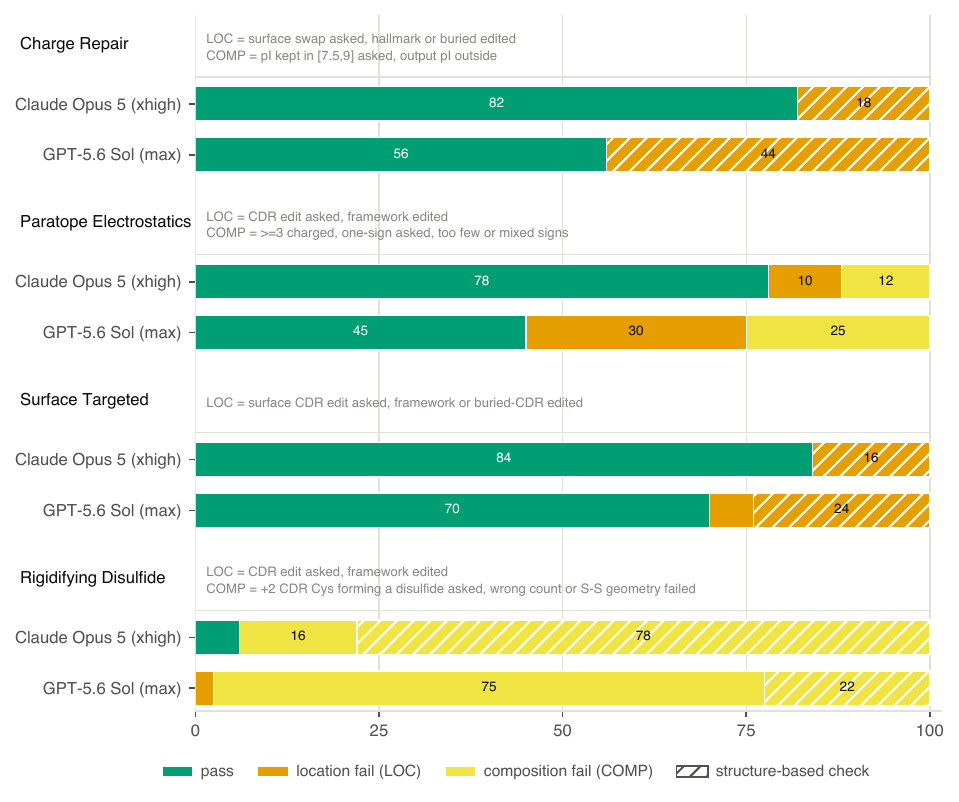}
\caption{\textbf{Without tools, failures split into region errors and
count errors.} Instruction-following failure
modes for the four most challenging tasks without tools, shown
for Opus 5 (xhigh) and GPT-5.6 sol. Each cell is 50 outputs (5 seeds
$\times$ 10 outputs). \emph{location} (LOC) failures edit
outside the region the task names, while \emph{composition} (COMP) failures
edit the right region but miss its target counts/ranges; the grey text above each task gives the task-specific form of both.}
  \label{fig:if-failure-modes}
\end{figure}

For the four hardest tasks without tools or design history, all instruction-following failures fall into two must-have requirement categories (Figure~\ref{fig:if-failure-modes}). \emph{Location} errors edit outside the region the task
names, and \emph{composition} errors edit the right region but miss a target
count or range. Every failing criterion is one that the provided tools can measure directly, which is why exposing them lifts the strict pass rate to 1.00 in nine of twelve arms

\section{Oracle performance for black-box setting}\label{appendix:bbfigure}
Black-box agents are evaluated across six models and five seeds to assess the effect of tool access on oracle performance. 

\begin{figure}[h]
  \centering
  \includegraphics[width=0.55\textwidth]{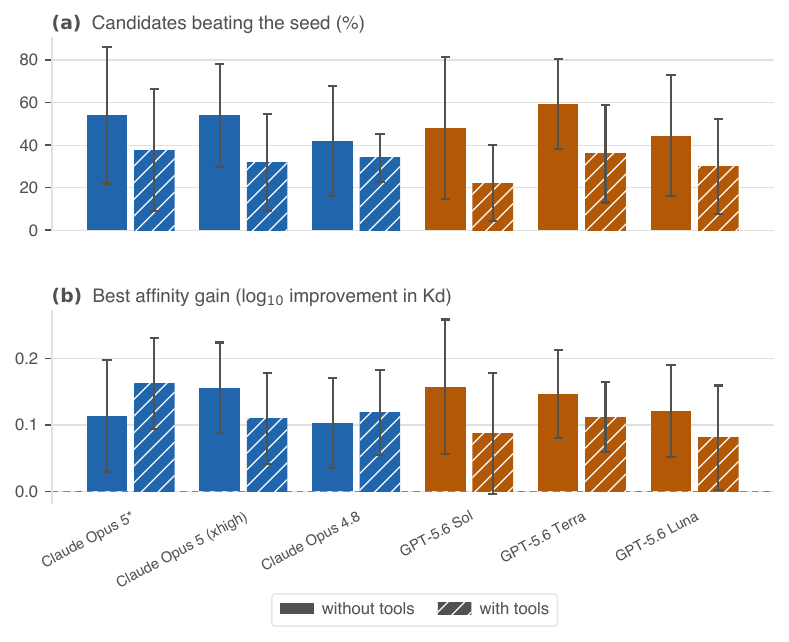}
\caption{\textbf{Black-box setting: tools do not help oracle performance.} The black-box agent run on 5 seeds, with 10 sequences generated per
seed. Bars are means over the 5 input seeds with $\pm1$ SD whiskers; solid bars
withhold tools, hatched bars provide them, and colour marks the model family
(Claude blue, GPT orange). \emph{Candidates beating the seed} is the fraction of outputs improving on the seed; \emph{best affinity gain} is the largest improvement among the 10 outputs. 
Effort is \texttt{max} unless noted. $^{*}$Claude Opus~5 (\texttt{max}) averages four seeds in the with-tools setting, as one seed resulted in a refusal.}\label{fig:black-box task}
\end{figure}

\end{document}